\documentclass[]{llncs}
\usepackage{amsmath}
\usepackage{graphicx}
\usepackage{tikz,pgfplots}
\usepackage{tabularx}
\usepackage{siunitx}

\newcommand{\etal}[1]{#1~\textit{et~al.}}

\title{Phase-Sensitive Region-of-Interest\\Computed Tomography}
\author{Lina Felsner\inst{1}	\and Martin Berger\inst{3} \and Sebastian Kaeppler\inst{1} \and Johannes Bopp\inst{1}  
	\and Veronika Ludwig\inst{2} \and Thomas Weber\inst{2} \and Georg Pelzer\inst{2} \and Thilo Michel\inst{2} \and Andreas Maier\inst{1} \and \\Gisela Anton\inst{2} \and Christian Riess\inst{1} }
\institute{Pattern Recognition Lab, Computer Science, Univ. of Erlangen-N{\"u}rnberg\\
	\and Erlangen Centre for Astroparticle Physics, Univ. of Erlangen-N{\"u}rnberg 
\and Siemens Healthcare GmbH}

\begin{document}

\maketitle

\begin{abstract}
X-Ray Phase-Contrast Imaging~(PCI) yields absorption, differential phase, and dark-field images. Computed Tomography (CT) of grating-based PCI can in principle provide high-resolution soft-tissue contrast. Recently, grating-based PCI took several hurdles towards clinical implementation by addressing, for example, acquisition speed, high X-ray energies, and system vibrations. However, a critical impediment in all grating-based systems lies in limits that constrain the grating diameter to few centimeters.

In this work, we propose a system and a reconstruction algorithm to circumvent this constraint in a clinically compatible way. We propose to perform a phase-sensitive Region-of-Interest (ROI) CT within a full-field absorption CT. The biggest advantage of this approach is that it allows to correct for phase truncation artifacts, and to obtain quantitative phase values. Our method is robust, and shows high-quality results on simulated data and on a biological mouse sample. This work is a proof of concept showing the potential to use PCI in CT on large specimen, such as humans, in clinical applications.
\end{abstract}

\section{Introduction}
\begin{figure}[bt]
	\centering
	\begin{tabular}{cccc}
		\includegraphics[height=0.2\textwidth]{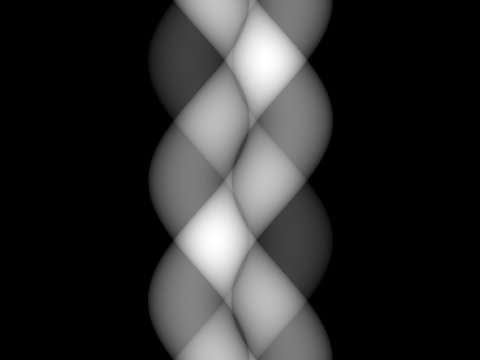} &
		\includegraphics[height=0.2\textwidth]{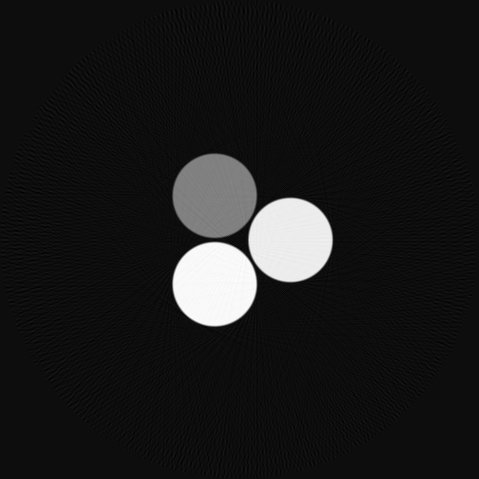} &
		\quad
		\includegraphics[height=0.2\textwidth]{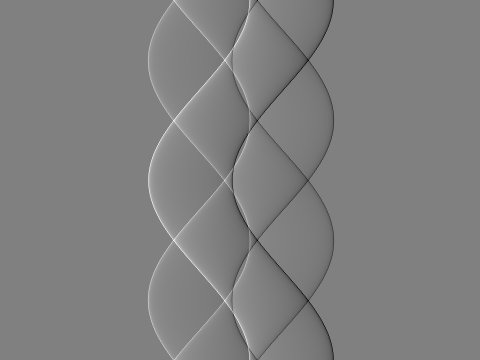} &
		\includegraphics[height=0.2\textwidth]{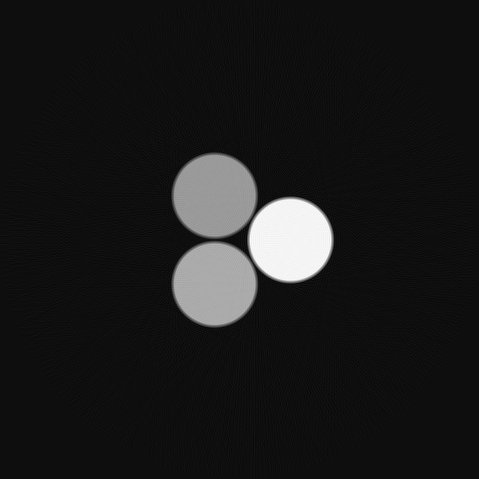}
	\end{tabular}
	\caption{Sinogram and reconstruction of the absorption (left) and differential phase images (right) of three cylinders with different materials at 82~keV. From the top in clockwise direction: water, PTFE and PVC.}
	\label{fig:simulations}
\end{figure}

X-ray Phase-Contrast Imaging (PCI) is a novel imaging technique that can be
implemented with an X-ray grating interferometer~\cite{Pfeiffer06-PRA}.
Such an interferometer provides an X-ray absorption image, and additionally a
differential phase-contrast image and a dark-field image.
X-ray absorption
and phase encode material-specific parameters that are linked
to the complex index of refraction~$n$, given as
$n = 1- \delta + \mathrm{i} \cdot \beta$. Here, $\delta$ relates to the
phase shift and $\beta$ to the attenuation. Since PCI yields high soft tissue contrast~\cite{Donath10:TCX,Koehler15:SDX}, it is particularly interesting to apply it in Computed Tomography~(CT).
Figure~\ref{fig:simulations} shows example sinograms for the absorption and phase, and the associated tomographic reconstructions. It furthermore shows that their information is exclusive, allowing for the distinction of different materials.

One key advantage of a grating-based interferometer is its compatibility with
clinical X-ray equipment~\cite{Pfeiffer06-PRA}. 
For clinical application, several practical challenges were recently addressed. Among these works are significant improvements in acquisition speed~\cite{Bevins12:MXC}, higher X-ray energies to penetrate large bodies~\cite{Gromann17}, and system vibrations~\cite{Horn18-IOA}.

However, one major obstacle to clinical implementation lies in the fact that
it is not clear whether gratings with a diameter of more than a few centimeters can be integrated in a clinical CT system, due to increasing vibrations sensitivity, production cost and complexity for larger grating sizes.
Such a small field of view is a major challenge for medical applications, since it only allows to reconstruct a small region of interest (ROI). This leads to difficult region localization and limited information on surrounding tissue.
Furthermore, the object is typically larger than the field of view, which leads to truncation in the projection images.

Truncation is a substantial issue in the reconstruction of conventional projection images, 
leading to artifacts, such as cupping.
\etal{Noo} showed that the reconstruction of a ROI from truncated differential images can be accurately obtained in certain cases~\cite{Noo04-ATH}. Unfortunately, for the so-called interior problem, where the ROI is completely inside the object, there is no unique solution.
\etal{Kudo}~\cite{Kudo08-TAP} found later, that the solution is unique if prior knowledge on the object is available in the form, that the object is known within a small region located inside the region of interest.
For differential projection data, the interior problem can be approached iteratively~\cite{Cong12-DPC,Lauzier12-ITI}.
However, these approaches are of limited practical use due to strong assumptions or high computational demand.

In this work, we propose a methodology for phase-sensitive region-of-interest
imaging within a standard CT. The key idea is to solve the shortcomings of
existing ROI imaging by complementing the small-area phase measurements with
the full-field absorption signal, which is similar in spirit to the work by \etal{Kolditz}~\cite{Kolditz10:VOI}. To this end, we propose to mount a
grating-based system in the center of an absorption CT system. The truncated
phase signal can be extrapolated beyond the grating limits using the full
absorption information and the phase within the ROI. This mitigates the typical
truncation artifacts, and even provides quantitative phase information within
the ROI, thereby paving the way towards phase CT in a clinical environment.

\section{Methods}
The proposed method consists of a system and a reconstruction algorithm. We describe the system in Sec.~\ref{sub:setup} and the algorithm in Sec.~\ref{sub:truncation}.
	
\subsection{Realization of the System}\label{sub:setup}

\begin{figure}[tb]
	\centering
	\def\svgwidth{\linewidth}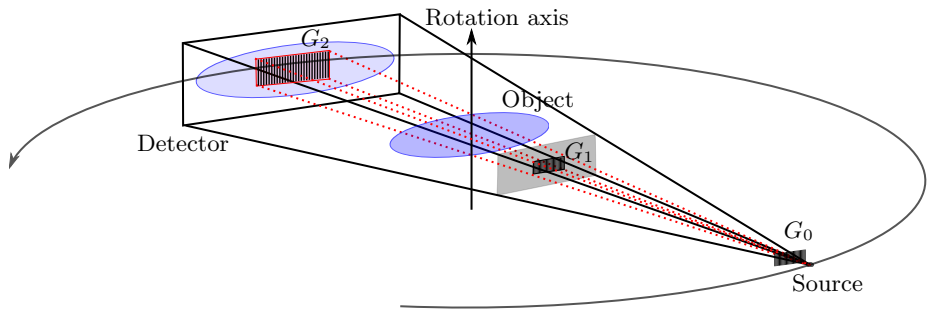
	\caption{Setup of the proposed imaging system.}
	\label{fig:setup}
\end{figure}

A grating-based (Talbot-Lau) interferometer consists of three gratings $G_0$, $G_1$, $G_2$ that are
placed between X-ray tube and detector (see Fig.~\ref{fig:setup}). $G_0$ is placed close to the source to
ensure spatial coherence.  $G_1$ is located in front of the object to
imprint a periodical phase shift onto the wave front.  $G_2$ is located in
front of the detector to resolve sub-resolution wave modulations.

Recently, the implementation of an interferometer into a clinical-like C-arm setup was demonstrated \cite{Horn18-IOA}.
We propose an embedding of the gratings in a clinical imaging system, such that the gratings cover only a region of interest.
An attenuating collimator can be used for mounting the gratings, leading to less dose in the Peripheral Region (PR) outside of the gratings. 
PCI has a dose advantage compared to attenuation for high-resolution detectors~\cite{Raupach11-AEO}, which suggests that it could also be advantageous to perform a high-resolution reconstruction in the ROI and reduce the resolution outside of the grating area to save dose.
Figure~\ref{fig:setup} shows a sketch of the setup.

While the geometry of the full absorption image is a cone-beam, the smaller grating area exhibits approximately parallel beams, simplifying reconstruction. For example, with a source-detector distance of \SI{200}{\centi\meter} and a grating size of \SI{5}{\centi\meter}, the maximal angle is \mbox{$\sim1.5^\circ$}. We apply a RamLak filter to the truncation-free absorption signal and a Hilbert filter to the phase-signal.

\subsection{Truncation Correction}\label{sub:truncation}
\begin{figure}[tb]
	\centering
	\def\svgwidth{\linewidth}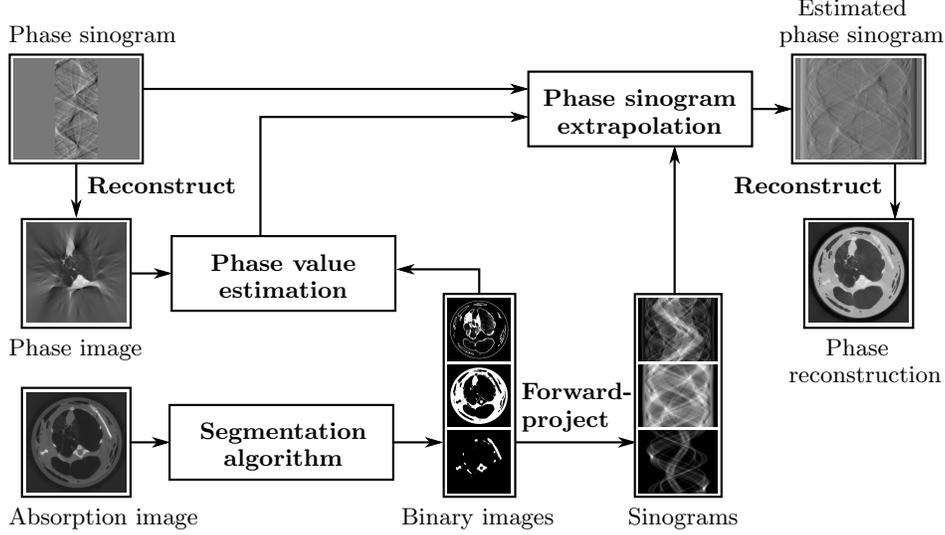
	\caption{Overview of the proposed method. A segmentation of the materials allows to obtain an estimate of their respective phase values. A non-truncated sinogram is extrapolated from the truncated sinogram and the extrapolated phase values.}
	\label{fig:pipeline}
\end{figure}

The pipeline of the proposed algorithm is shown in Fig.~\ref{fig:pipeline}.  We
first perform a reconstruction of absorption and truncated phase, and segment
the absorption into $k$ materials. This allows to estimate the phase values per
material within the ROI, and to extrapolate the phase values across the full
area.
Reconstructing now the phase signal from the phase measurements within the ROI
and the extrapolated phase values outside, we obtain a non-truncated phase
reconstruction.

\textbf{Segmentation Algorithm.} 
The absorption signal is decomposed into $k$ materials via segmentation. While in principle any algorithm could be used here, we fitted a Gaussian Mixture Model with $k$ components to the histogram.

\textbf{Phase Value Estimation.}
The truncated phase ROI is reconstructed. 
For each material, we estimate its phase value $\delta_k$ by computing the mean over its segmented pixels in the ROI. If a material is not contained in the ROI, we heuristically set it to the mean $\delta$ over the ROI. 
Note that the estimated values will be differentiated in the subsequent processing which effectively removes any bias from the estimated values.

\textbf{Phase Sinogram Extrapolation.}
The phase shift is a line integral of delta coefficients,
\begin{equation}
\phi = \int \delta \, \textup{d} z = \sum_k \left( \delta_k \cdot \int_k \textup{d} z \right)\enspace ,
\label{eq:phaseShift}
\end{equation}
that can be decomposed into $k$ materials. 
We insert Eqn.~\ref{eq:phaseShift} into the measured differential phase signal,
\begin{equation}
\varphi = \frac{\lambda \cdot d}{2\pi \cdot p_2} \frac{\partial \phi}{\partial x} = \frac{\lambda \cdot d}{2\pi \cdot p_2}  \sum_k \left( \frac{\partial \; \delta_k \int_k  \textup{d} z}{\partial x} \right)\enspace ,
\end{equation}
consisting of sensitivity direction $x$, wavelength $\lambda$, $G_1$-$G_2$
distance $d$, and the \mbox{$G_2$ period} $p_2$.
The factor $(\lambda \cdot d) / (2 \pi \cdot p_2)$ is the setup sensitivity,
which is a material-independent scaling factor and can therefore be ignored.
Thus, we can obtain the sinogram of the differential phase by applying the
derivative to the $\delta_k$-weighted line integrals, given by the
forward projections of the segmented materials. 
That way, the truncated phase sinogram is extrapolated with the missing
sinogram information outside the ROI. 
In our implementation, we anchored the phase-shift of air to zero.
The extrapolated phase together with the measured ROI allow for a quantitative phase
reconstruction.

\section{Experiments}
We evaluate our approach on a biological sample and on simulated data. We compare the results in both cases with a ground-truth reconstruction of the untruncated data, where we applied a 3~$\times$~3~pixel smoothing with a median filter.
The line plots are obtained in horizontal and vertical direction through the
center of the reconstructed ROI. The quantitative metrics are the Root
Mean Square Error~(RMSE) and the Structural Similarity~(SSIM) inside the region of interest.

\begin{figure}[tb]
	\centering
	\begin{tabular}{ccc}
		\includegraphics[width=0.3\textwidth]{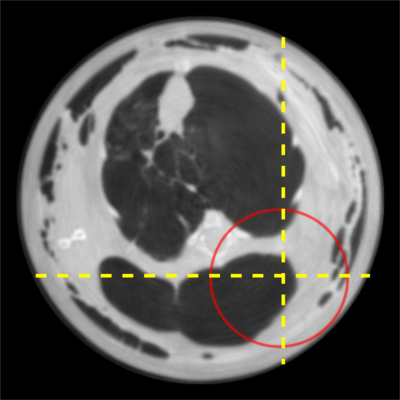} & 
		\quad
		\includegraphics[width=0.3\textwidth]{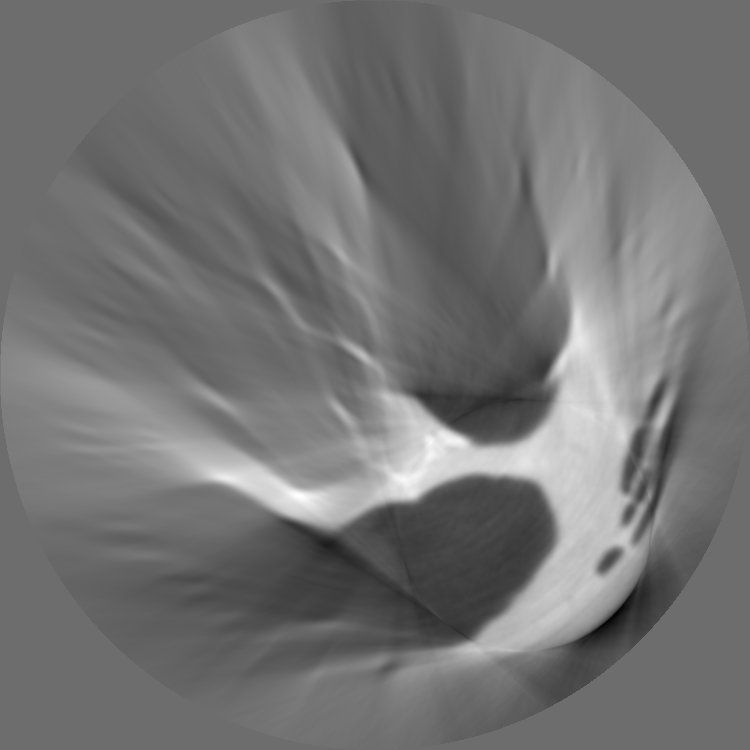} & %
		\quad
		\includegraphics[width=0.3\textwidth]{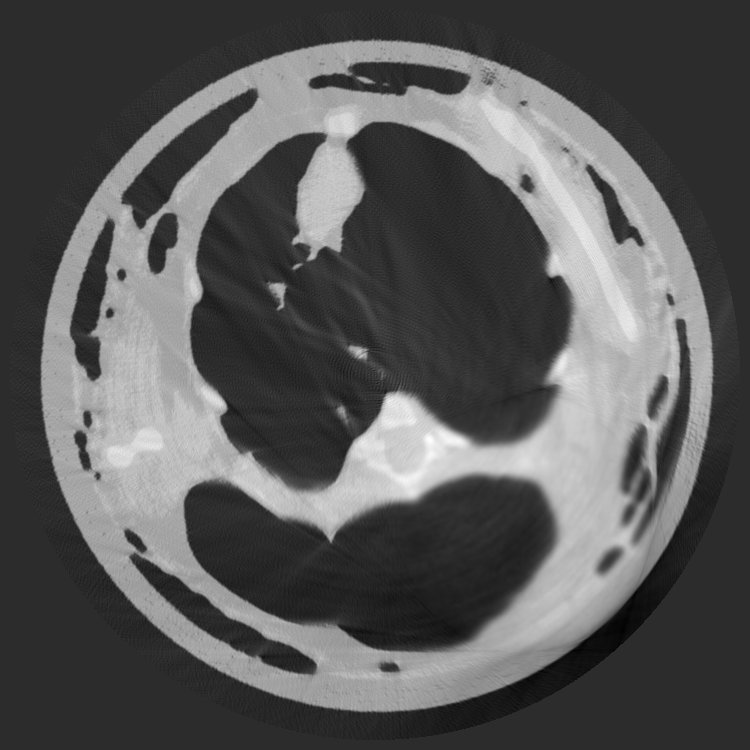}
	\end{tabular}
	\caption{Mouse sample. Left: ground truth with ROI (red) and line profile (yellow). Center: truncated phase reconstruction. Right: proposed phase reconstruction.} 
	\label{fig:estimation_ROI8}
\end{figure}

\begin{figure}[tb]
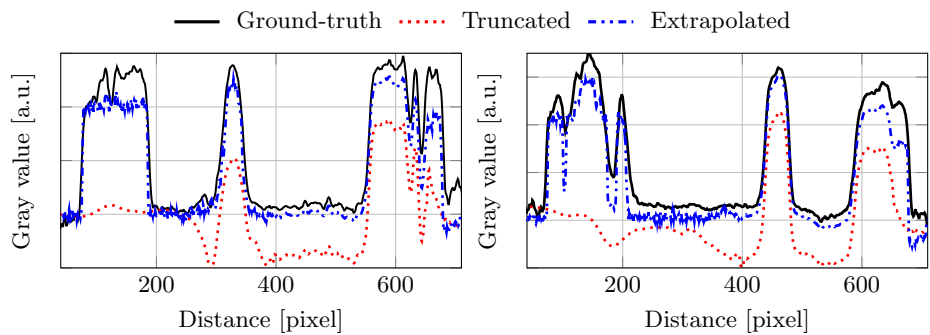

	\centering
	    \newenvironment{customlegend}[1][]{%
        \begingroup
        \csname pgfplots@init@cleared@structures\endcsname
        \pgfplotsset{#1}%
    }{%
        \csname pgfplots@createlegend\endcsname
        \endgroup
    }%
    \def\addlegendimage{\csname pgfplots@addlegendimage\endcsname}

\begin{tikzpicture}
\begin{customlegend}[legend columns=5,legend style={align=left,draw=none,column sep=1ex},
        legend entries={Ground-truth, Truncated, Extrapolated, Corrected}]
        \addlegendimage{color=black,line width=1.2pt,solid}
	\addlegendimage{color=red, line width=1.2pt, dotted}
        \addlegendimage{color=blue, line width=1.2pt, dash dot dot}  
        \end{customlegend}
\end{tikzpicture}
	\hspace*{-0.5cm}
	\begin{tabular}{cc}
		\resizebox{0.5\textwidth}{!}{\input{mouse_line_plot_x.tikz}}&
		\resizebox{0.5\textwidth}{!}{\input{mouse_line_plot_y.tikz}}
	\end{tabular}
	\vspace*{-0.3cm}
	\caption{Line plots of the mouse sample through the region of interest. Left: in horizontal direction; Right: in vertical direction.} 
	\label{fig:line_plots_ROI8}
\end{figure}

\subsection{Biological Sample}\label{sub:mouse}
We use a scan of a mouse \cite{Weber12-IOT} 
as biological sample with complex anatomical structures. The scan is
manually truncated by cropping the ROI in the sinogram. 
This allows us to compare the results to a full reconstruction.
The acquisition setup consists of a tungsten anode X-ray tube at 60\,kVp and a
Varian PaxScan~2520D~detector with \SI{127}{\micro\meter} pixel pitch. The
grating periods are \SI{23.95}{\micro\meter}, \SI{4.37}{\micro\meter} and
\SI{2.40}{\micro\meter} for $G_0$, $G_1$, and $G_2$, respectively.
The $G_0-G_1$~distance is \SI{161.2}{\centi\meter}. Acquisition is done with 8 phase steps with exposures of 3.3\,s each and a tube current of 30\,mA.
The image sequence contains 601~projection images over a full circle.
We chose a ROI size of a third of the detector size.

We evaluate our method for nine different ROIs on the mouse, with $k$
empirically set to $5$. Our algorithm successfully reduces the truncation
artifacts for all ROIs.  Figure~\ref{fig:estimation_ROI8} shows example
reconstructions of ground truth, truncated phase and estimated phase.
The benefit of the truncation correction can be recognized both inside and outside of the ROI. The surrounding tissue of the ROI exhibits slightly sharpened edges due to the segmentation boundaries.
The line plots after extrapolation in Fig.~\ref{fig:line_plots_ROI8} are close to the ground truth.
Table~\ref{tab:quant} depicts RMSE and SSIM relative to the ground-truth. The proposed algorithm reduces the RMSE by more than 50\,\%, and SSIM by 64\,\%.

\begin{figure}[tb]
	\centering
	\begin{tabular}{cccc}
			\includegraphics[width=0.3\textwidth]{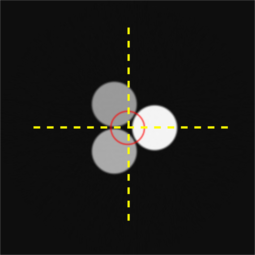} & 
			\quad
			\includegraphics[width=0.3\textwidth]{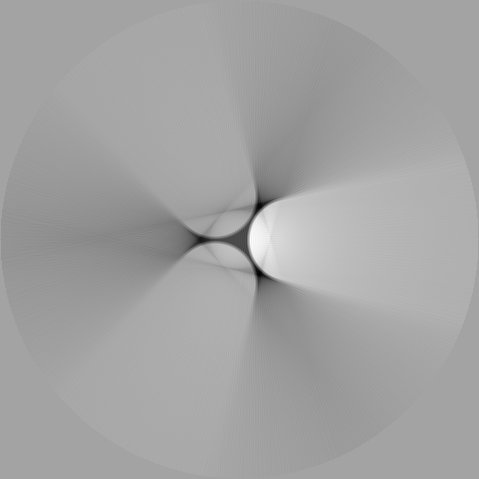} & %
			\quad
			\includegraphics[width=0.3\textwidth]{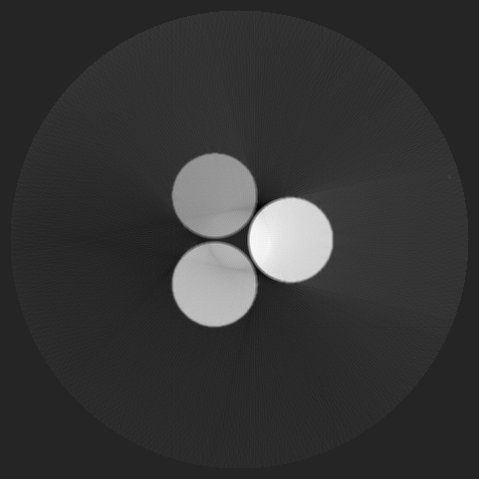}
	\end{tabular}
	\caption{Simulated sample. Left: ground truth with ROI (red) and line
	profile (yellow).  Center: truncated phase reconstruction. Right: proposed
	phase reconstruction.}
	\label{fig:estimation_sim}
\end{figure}

\begin{figure}[tb]
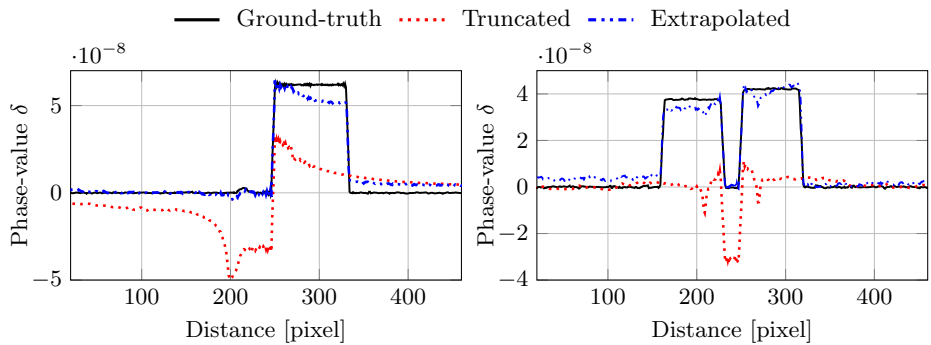

	\centering
	    \newenvironment{customlegend}[1][]{%
        \begingroup
        \csname pgfplots@init@cleared@structures\endcsname
        \pgfplotsset{#1}%
    }{%
        \csname pgfplots@createlegend\endcsname
        \endgroup
    }%
    \def\addlegendimage{\csname pgfplots@addlegendimage\endcsname}

\begin{tikzpicture}
\begin{customlegend}[legend columns=5,legend style={align=left,draw=none,column sep=1ex},
        legend entries={Ground-truth, Truncated, Extrapolated, Corrected}]
        \addlegendimage{color=black,line width=1.2pt,solid}
	\addlegendimage{color=red, line width=1.2pt, dotted}
        \addlegendimage{color=blue, line width=1.2pt, dash dot dot}  
        \end{customlegend}
\end{tikzpicture}
	\hspace*{-0.5cm}
	\begin{tabular}{cc}
		\resizebox{0.5\textwidth}{!}{\input{sim_line_plot_x.tikz}}&
		\resizebox{0.5\textwidth}{!}{\input{sim_line_plot_y.tikz}}
	\end{tabular}
	\vspace*{-0.3cm}
	\caption{Line plots of the simulated sample at 82~keV. Left: horizontal direction. Right: vertical direction.}
	\label{fig:line_plots_sim}
\end{figure}

\subsection{Quantitative Evaluation}\label{sub:qunatitaitve}
The quantitative data was created by simulations using the phase material
values for water, polyvinylchlorid~(PVC), and polytetrafluorethylen~(PTFE) at
82~keV from the literature~\cite{Willner13-QXR}.  We used a parallel beam
geometry and 360 projection images over a full circle. The size of the ROI is
set to $1/8$ of the detector width, $k$ is set to $4$.

The absorption and phase reconstructions are shown in Fig.~\ref{fig:simulations}.
As PTFE and PVC have very similar absorption values, the segmentation erroneously labels them as identical materials. 
However, the proposed approach is robust to such a missegmentation, which can
be recognized by the well distinguishable phase values of PVC and PTFE in
Fig.~\ref{fig:estimation_sim}.  
This is supported by the line plots in Fig.~\ref{fig:line_plots_sim}, where the correctness of the quantitative phase values can also be verified.
The measurements in Tab.~\ref{tab:quant} support the visual impression, with an average improvement of over 50\,\% for the SSIM and one magnitude decrease for the RMSE.
Unfortunately, the phase value of PTFE in Fig.~\ref{fig:line_plots_sim} (left)
decreases slowly outside the ROI with increasing distance to the ROI.

\begin{table}[tb]
	\centering
	\caption{Quality metrics with respect to the ground truth inside of the ROIs. 
	Mean and standard deviation over 16 ROIs (simulated data) and 9 ROIs (mouse data).}
	\begin{tabularx}{\textwidth}{lXcXcXcXc}
		\hline\noalign{\smallskip}
		&& \multicolumn{3}{c}{Simulation} && \multicolumn{3}{c}{Mouse} \\
		&& RMSE && SSIM && RMSE && SSIM \\
		\noalign{\smallskip}
		\hline
		\noalign{\smallskip}
		Truncated 	&& $2.88E-08 \pm 6.59E-09 $ 	&& $0.46 \pm 0.30 $ && $ 2.08 \pm 0.27 $  &&	$0.30 \pm 0.33 $	\\ 
		Estimated 	&& $3.00E-09 \pm 1.17E-09 $ 	&& $0.99 \pm 0.00 $ && $ 0.66 \pm 0.14$   && $ 0.94 \pm 	0.05$ \\
		\hline
	\end{tabularx}
	\label{tab:quant}
\end{table}

\begin{figure}[tb]
	\hspace*{-0.5cm}
	\begin{tabular}{cc}
		\resizebox{0.5\textwidth}{!}{\begin{tikzpicture}
\begin{axis}[
width=\textwidth,
height=\axisdefaultheight,
scale = 0.5,
xlabel={ROI [\% of the dectector width]},
xlabel near ticks,
xticklabels={0,0,1/4,1/6,1/8,1/10},
ylabel={RSME},
ymin=0,
ymax=4E-8,
y label style={at={(axis description cs:0.1,.5)}},
ymajorgrids=true,
]
\addplot [color=red, only marks,mark=o,]
plot [error bars/.cd, y dir = both, y explicit]
table[row sep=crcr, y error index=2]{	
	1 1.90604140142966E-08 1.17764033529389E-09\\
	2 2.85E-08 2.64895185180288E-09\\
	3 3.26E-08 2.74825704947246E-09\\
	4 3.50E-08 2.2335242592868E-09\\
	};
\addplot [color=blue, only marks,mark=o,]
plot [error bars/.cd, y dir = both, y explicit]
table[row sep=crcr, y error index=2]{	
	1 2.79E-09 7.99570457164346E-10\\
	2 2.69E-09 0.000000001\\
	3 3.15E-09 1.50513586467508E-09\\
	4 3.38E-09 1.6192380722267E-09\\
	};
\end{axis}
\end{tikzpicture}%
}&
		\resizebox{0.5\textwidth}{!}{\begin{tikzpicture}
\begin{axis}[
width=\textwidth,
height=\axisdefaultheight,
scale = 0.5,
xlabel={ROI [\% of the dectector width]},
xlabel near ticks,
ylabel={SSIM},
xticklabels={0,0,1/4,1/6,1/8,1/10},
y label style={at={(axis description cs:0.1,.5)}},
ymajorgrids=true,
legend pos=south west,
]
\addplot [color=red, only marks,mark=o,]
plot [error bars/.cd, y dir = both, y explicit]
table[row sep=crcr, y error index=2]{	
	1 7.97E-01 0.0202701794 \\
	2 5.42E-01 0.0888353138 \\
	3 3.54E-01 0.2148996644 \\
	4 1.81E-01 0.3575979844 \\
};
\addlegendentry{Truncated}
\addplot [color=blue, only marks,mark=o,]
plot [error bars/.cd, y dir = both, y explicit]
table[row sep=crcr, y error index=2]{	
	1 9.95E-01 0.0018097127 \\
	2 9.95E-01 0.0009069817 \\
	3 9.94E-01 0.0024511833 \\
	4 9.94E-01 0.0026609789 \\
};
\addlegendentry{Estimated}
\end{axis}
\end{tikzpicture}%
}
	\end{tabular}
	\caption{Performance for simulated data at 82\,keV, averaged over different ROI positions for 4 different ROI sizes. Red: truncated reconstruction. Blue: estimated reconstruction.}
	\label{fig:box_plot}
\end{figure}
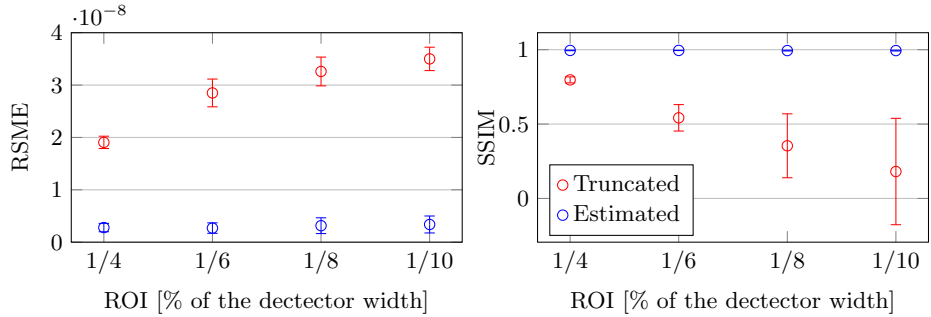

We also investigate the influence of the ROI size at four different locations, pushing the ROI away from the center. The mean error and standard deviation for RMSE and SSIM are visualized in Fig.~\ref{fig:box_plot}.
The quality of the truncated reconstruction is significantly decreased by a smaller ROI.
Contrary, the proposed method is remarkably robust to changes in size and
location of the ROI. 

\section{Conclusion}
We propose a system and a method to perform quantitative ROI reconstruction of
phase CT. The idea is to embed a grating interferometer into a standard CT, and
to extrapolate the phase beyond the ROI with the absorption
information to reduce truncation artifacts.
Our results on quantitative data and a real biological sample
are highly encouraging, and we believe that this is an important
step towards using PCI on a clinical setup for larger samples.

\subsubsection*{Acknowledgments.}
Lina Felsner is supported by the  International Max Planck Research School - Physics of Light (IMPRS-PL). 
\subsubsection*{Disclaimer.}
The concepts and information presented in this paper are based on research and are not commercially available. 

\bibliographystyle{splncs}
\bibliography{lit}

\end{document}